\documentclass[twocolumn,secnumarabic,amssymb, nobibnotes,superscriptaddress,showpacs, prb]{revtex4-1}
\usepackage[utf8]{inputenc}
\usepackage{amsmath}
\usepackage{color}
\usepackage{graphicx}
\usepackage{subfigure}
\usepackage{multirow}
\usepackage{float}
\usepackage{bm}

\date{\today}

\begin{document}

\title{Impact of domains on the orthorhombic-tetragonal 
transition of BaTiO$_3$: \newline an {\it{ab initio}} study}
\author{Anna Gr\"unebohm}
\affiliation{ICAMS, Ruhr-University Bochum, Bochum, Germany}
\author{Madhura Marathe}
\affiliation{Institut de Ci\`encia de Materials de Barcelona (ICMAB-CSIC), 08193 Bellaterra, Spain.}
\affiliation{Universitat Autònoma de Barcelona, 08193 Bellaterra, Spain.}
\begin{abstract}
 We investigate the multi-domain structures in the tetragonal and 
 orthorhombic phases of BaTiO$_3$ and the impact of the presence 
 of domain walls on the intermediary phase transition. We focus on 
 the change in the transition temperatures resulting from various 
 types of domain walls and their coupling with an external electric 
 field. We employ molecular dynamics simulations of an {\it{ab initio}} 
 effective Hamiltonian in this study. After confirming that this model 
 is applicable to multi-domain configurations, we show that the phase 
 transition temperatures strongly depend on the presence of  
 domains walls. Notably we show that elastic 90$^{\circ}$ walls can 
 strongly reduce thermal hysteresis. 
 Further analysis shows that the change in transition temperatures 
 can be attributed to two main factors -- long-range monoclinic 
 distortions induced by walls within domains and domain wall widths.  
 We also show that the coupling with the field further facilitates  
 the reduction of thermal hysteresis for orthorhombic 90$^{\circ}$ 
 walls making this configuration attractive for future applications. 
\end{abstract}
\maketitle

\section{Introduction}
\label{sec:intro}
In the last decades the functional properties of ferroelectric (FE) 
perovskites came into focus for exciting 
applications.~\cite{Scott_2007} 
In particular Pb-free BaTiO$_3$-based materials are promising for 
sustainable technologies with operating temperatures at and below 
room temperature.~\cite{Acosta_et_al_2017} BaTiO$_3$ (BTO) 
crystallizes in the paraelectric (PE) cubic phase ($Rm\bar{3}m$) at 
high temperatures and below 120$^{\circ}$C, 5$^{\circ}$C and 
-80$^{\circ}$C the FE tetragonal (T), orthorhombic (O) and 
rhombohedral (R) phases with $P4mm$, 
$Amm2$ and $R3m$ symmetry 
occur.~\cite{Lines-Glass,Limboek_Soergel_2014,Hippel_1950}
For applications close to ambient temperatures thus in particular 
the T and O ferroelectric phases and the transition between them 
are relevant. 
There are only a few reports which have studied the T-O transition in 
detail.~\cite{Limboek_Soergel_2014,Hippel_1950,Doering_Eng_Kehr_2016,Eisenschmidt_et_al_2012,hershkovitz_mesoscopic_2020} 
X-ray diffraction measurements have pointed to the existence of 
an intermediate phase with short-range monoclinic 
order at this transition~\cite{Eisenschmidt_et_al_2012,kalyani_metastable_2015} 
and 
under an applied field the T-phase transforms to a monoclinic phase 
which remains stable after removing the field.~\cite{Cao_et_al_2009} 
Another measurement also confirmed the existence of monoclinic phases 
and reported a ``thermotropic'' phase 
boundary~\cite{Lummen_et_al_2014} which is the temperature-analog of 
concentration-dependent morphotropic phase boundaries. 
Such phase boundaries are related to  flat energy landscapes, 
enhanced polarization rotation and in turn exceptional large 
functional  responses.~\cite{Guo_et_al_2000,Fu_Cohen_2000}

The T--O transition is a first-order phase transition with phase 
coexistence and a 
broad thermal hysteresis between the transition temperatures for 
cooling ($T_C^c$) and heating ($T_C^h$).~\cite{Lines-Glass} 
Furthermore, different nature of the phase transition with abrupt 
versus continuous changes 
in polarization while cooling and heating has been reported.~\cite{Limboek_Soergel_2014,Bai_et_al_2017} 
The measured transition temperatures for the T--O transition have a 
large error bar of more than 10\,K which is comparable to the 
magnitude of the hysteresis itself.~\cite{Limboek_Soergel_2014,Hippel_1950,Doering_Eng_Kehr_2016,Eisenschmidt_et_al_2012,Bai_et_al_2017}
Hysteresis potentially results in reduced reversible 
functional responses and a wide range of responses for  
different samples, making the material 
unsuitable for devices which require long operational 
lifetime.~\cite{Marathe_PSS_2018,Stern-Taulats_et_al_2016}  
Therefore, it is important to understand the 
extrinsic and intrinsic factors which contribute to 
thermal hysteresis and the order of the transition. 
So far, the large error bars in  $T_C$'s have been attributed to 
presence of domains.~\cite{Limboek_Soergel_2014,Bai_et_al_2017} 
Domains and domain walls are also known as one of the major 
sources of extrinsic functional responses.\cite{pramanick_domains_2012} 
Furthermore, recent developments in the field of sample 
preparation and measurement techniques with atomic resolution 
brings the idea of engineered domain walls (DW) within our reach  
and allows us to master hysteresis by microstructure design. 
Domain engineering designed for specific 
applications~\cite{Fousek_Cross_Ferro2003,Martin_Rappe_2016}  
and optimization of functional 
responses~\cite{Wada_et_al_2005,Rao_Wang_2007,Hlinka_et_al_2009}
have already been tested in experiments. 
Therefore, in the present study, we focus on domains and how 
the presence of multi-domain (MD) phases affects $T_C$, the nature 
of phase transition and its thermal hysteresis.

As dictated by underlying crystal symmetry, polarization $\bm{P}$ along 
all $\langle 100\rangle$ and $\langle 110\rangle$ directions would 
lead to degenerate states for T and O phase, respectively. 
The single-domain (SD) system with polarization along any one of 
these directions would be the ground state for ideal materials under 
perfect screening. 
However, typically multi-domain phases are 
prevalent in which domains of different polarization directions 
coexist and are separated by a domain wall in which the polarization 
changes direction. These walls are characterized by the angle of 
the polarization rotation along the domain wall normal, e.g., 
180$^{\circ}$ between domains with $\pm \bm{P}$ along $[010]$ or $[011]$, respectively.
Possible reasons for the stability of domain walls include local 
strain,\cite{Grunebohm_APL_2015} presence of defects\cite{ren_large_2004} and depolarization fields.\cite{kalinin_surface-screening_2018} Furthermore domains may nucleate in the course of the 
heat treatment which are at least meta-stable afterwards.~\cite{Wada_et_al_2005,noauthor_heterogeneous_2018} 

\begin{figure}[tb]
    \centering
\includegraphics[width=0.5\textwidth,clip,trim=1cm 10cm 16cm 0.5cm]{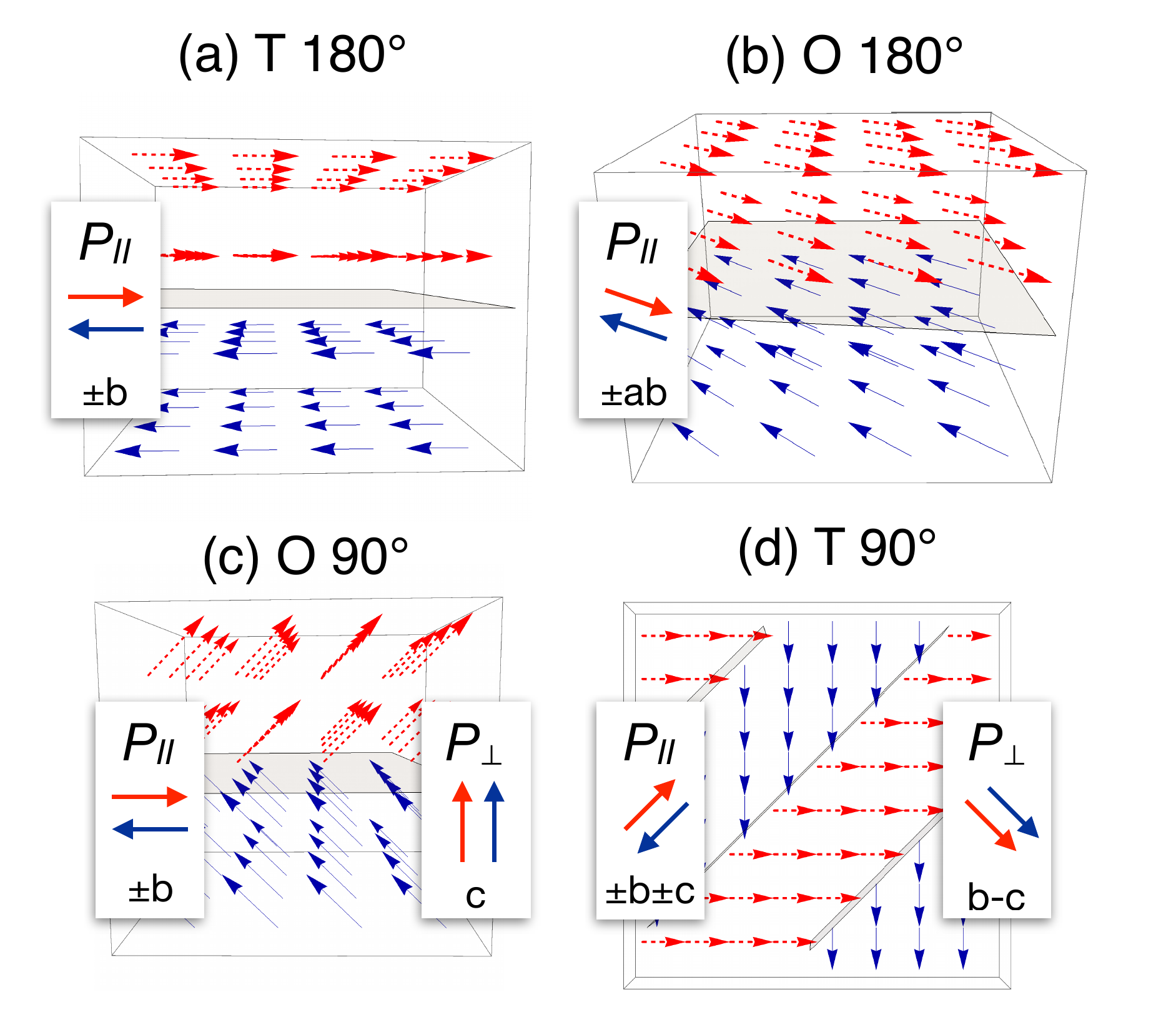}
    \caption{Local dipole configurations of the T and O phases of 
	BaTiO$_3$ under study:  (a)--(b) non-elastic 180$^{\circ}$ DWs, (c)--(d) 
	elastic 90$^{\circ}$ DWs. 
	Insets illustrate the polarization component parallel to the wall 
	($P_{||}$) along [010], i.e. along $\pm$b (a), [110],  (b),  (c) and [110] (d) 
	and the component perpendicular to elastic walls ($P_{\perp}$) 
	along [001], i.e. along c (c) and   [110] (d). For 180$^{\circ}$ walls, $P_{\perp}$ 
	is along [001].
    }
    \label{fig:domains}
\end{figure}

So far the understanding of the coupling of domain structure and 
the T--O transition is only rudimentary. 
Piezoresponse force 
microscopy has been used to image domain patterns within the 
T and O phases as well as evolution of domains at the phase 
transition.~\cite{Limboek_Soergel_2014,Doering_Eng_Kehr_2016} 
In general, it is however challenging to visualize and interpret 
the complex domain structures especially near phase transitions and disentangle the 
impact of superposition of various DW and their 
interactions, defects and inhomogeneities in the real sample. 
Theoretical simulations provide an ideal tool to 
separate the different elements of the microstructure by considering 
an ``idealised" system. 
However, previous theoretical studies have focused on the domain structure 
of the T or O phase either using density functional theory 
simulations at $0$\,K~\cite{Grunebohm_Gruner_Entel_2012,padilla_first-principles_1996} 
or phenomenological models~\cite{Marton_et_al_2010,Hlinka_2016}
away from the phase transition. 
Moreover, when the phase transition has been studied the impact of 
domains is usually neglected.~\cite{Bell:2001,Marathe_et_al_2017}

In the present paper, we close this gap by a systematic study of 
the T-O transition in the presence of domains focusing on specific 
low-energy configurations 
which are most common in experimental studies, namely planar 
charge-neutral 90$^{\circ}$ and 180$^{\circ}$ walls along high 
symmetry directions.~\cite{Limboek_Soergel_2014} 
In non-elastic 180$^{\circ}$ walls, the same strain occurs in 
adjacent domains (shown in Fig.~\ref{fig:domains}a-b),  
whereas the considered 90$^{\circ}$ DW (Fig.~\ref{fig:domains}c-d) 
are elastic. 
These walls are aligned along 
$\langle 100 \rangle$ and $\langle 110 \rangle$ in the O and T phases 
respectively. Note that the  alignment of 180$^{\circ}$ walls is not 
fully determined by symmetry, e.g., for $P_b$ and $P_{-b}$, all walls 
parallel to [001] are analogous to non-elastic charge-neutral walls,  
however, we focus only on the limiting cases being along the 
crystallographic $\langle100 \rangle$ directions;  
previous theoretical works\cite{Marton_et_al_2010,Lawless} have 
shown that these directions are favorable.  

For potential applications, not only thermal hysteresis and 
character of the transition, but also their coupling 
to an external field is relevant and would determine the functional 
properties. For example, it has been discussed in literature 
that giant caloric and piezoelectric responses are possible 
for a field-induced phase transition.\cite{Acosta_et_al_2017,moya_caloric_2014,Gruenebohm_review_2018,Hlinka_et_al_2009} 
Therefore, we study the effect of an applied field on the evolution 
of  multi-domain phases as a function of temperature as 
well as the density of domain walls. Our main aims are 
to understand the properties of MD phases around and at the T-O 
phase transition and to explore possibilities to engineer these 
phases to reduce the thermal hysteresis. 

The paper is organized as follows: the used {\it{ab initio}}-based 
method and simulation details are summarized in Sec.~\ref{sec:comp}, 
our results describing the properties of domain walls in T and 
O phases  and the interplay between domain structure and the 
phase transition are discussed in Sec.~\ref{sec:results}, 
and finally conclusions and outlook are given 
in Sec.~\ref{sec:conclude}. 

\section{Computational details}
\label{sec:comp}
\begin{figure}[tb]
    \centering
    \includegraphics[width=0.45\textwidth]{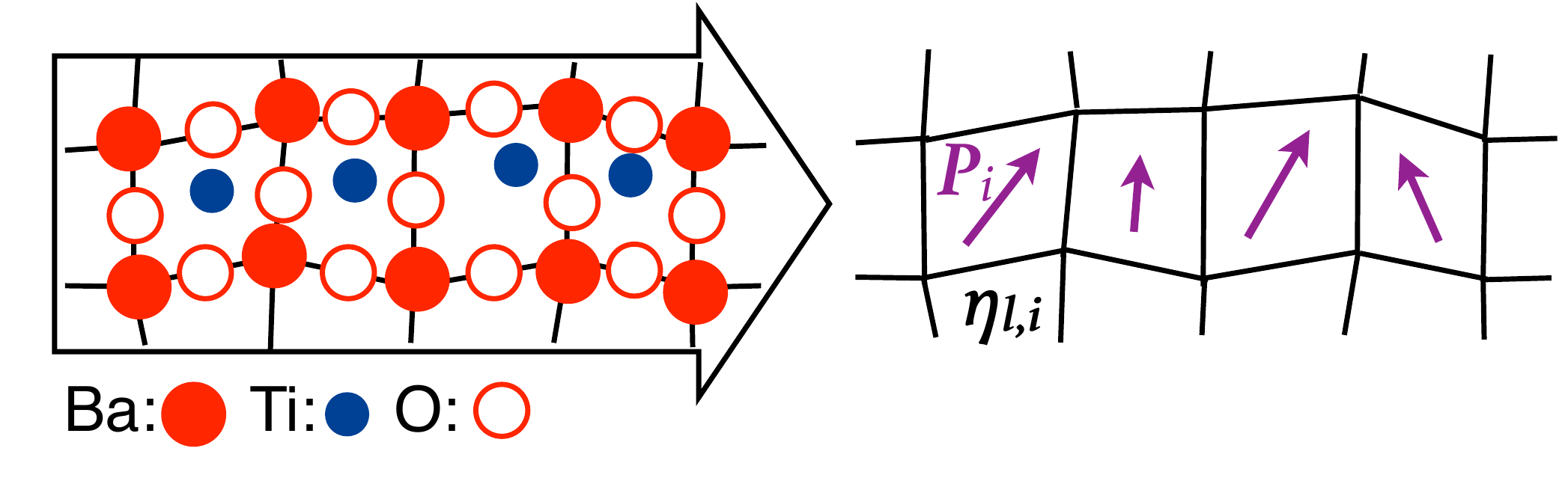}
    \caption{Illustration of the used coarse-grained model: 
	the electronic and atomic degrees of freedom are mapped 
	on the local dipole moment $\bm{u}_i$, i.e.\ the local 
	polarization $\bm{P}_i$ and the local strain $\bm{\eta}_{l,i}$.}
    \label{fig:model}
\end{figure}

We employ the effective 
Hamiltonian~\cite{Zhong_Vanderbilt_Rabe_1995,Zhong_Vanderbilt_Rabe_1994} 
defined for cubic perovskite ferroelectrics. Instead of treating all 
atomic positions as degrees of freedom, the collective atomic 
displacements are coarse-grained by local soft mode vectors $\bm{u}_i$
and local acoustic displacement vectors $\bm{w}_i$ of each unit 
cell $i$ in the simulation supercell. An internal optimization of  
$E(\bm{w}_i)$ is used reducing 
the degree of freedoms per formula unit from 5~atoms times the three 
Cartesian directions ($= 15$) to the three components of the soft mode 
$\bm{u}_i$  corresponding to the local polarization $\bm{P}_i$, 
see Fig.~\ref{fig:model}.  This method has been used extensively to 
study various properties 
such as phase diagrams,~\cite{BaStO,Marathe_et_al_2017,Kornev} 
domain structures\cite{Grunebohm_APL_2015,Feram2,Lai2} or functional 
responses~\cite{Marathe_et_al_2016,Ponomareva_Lisenkov_2012,Gui} of 
ferroelectric materials.

The effective 
Hamiltonian~\cite{Zhong_Vanderbilt_Rabe_1995,Zhong_Vanderbilt_Rabe_1994} 
is written as follows: 
\begin{eqnarray}
  \label{eq:EffectiveHamiltonian}
  \nonumber H^{\rm eff}
 & =& \frac{M^*_{\rm dipole}}{2} \sum_{i,\alpha}\dot{u}_{\alpha,i}^2\\
  &+& V^{\rm self}(\{\bm{u}\})+V^{\rm dpl}(\{\bm{u}\})+V^{\rm short}(\{\bm{u}\})\\
  \nonumber &+& V^{\rm elas,\,homo}(\eta_1,\dots\!,\eta_6)+V^{\rm elas,\,inho}(\{\bm{w}\})\\
  \nonumber &+& V^{\rm coup,\,homo}(\{\bm{u}\}, \eta_1,\cdots\!,\eta_6)+V^{\rm coup,\,inho}(\{\bm{u}\}, \{\bm{w}\}),
\end{eqnarray}
with $\eta_1,\dots,\eta_6$ being the six components of homogeneous strain 
in Voigt notation. 
The first term represents the kinetic energies of the local soft 
modes while  
$M^*_{\rm dipole}$ 
corresponds to the 
effective masses. 
Further, 
$V^{\rm self}(\{\bm{u}\})$ is the self-energy of the local mode, 
$V^{\rm dpl}(\{\bm{u}\})$ is the long-range dipole-dipole interaction, 
$V^{\rm short}(\{\bm{u}\})$ is the short-range interaction between 
local soft modes, 
$V^{\rm elas,\,homo}(\eta_1,\dots,\eta_6)$ is the elastic energy 
from homogeneous strains, 
$V^{\rm elas,\,inho}(\{\bm{w}\})$ is the elastic energy from 
inhomogeneous strains, 
$V^{\rm coup,\,homo}(\{\bm{u}\}, \eta_1,\dots,\eta_6)$ is the coupling 
between the local soft modes and the homogeneous strain, and 
$V^{\rm coup,\,inho}(\{\bm{u}\}, \{\bm{w}\})$ is the coupling between 
the soft modes and the inhomogeneous strains. 
The coupling with an external electric field $-Z^*E_i.u_i$ is  
also included in the Hamiltonian, where $Z^*$ is the Born effective charge 
associated with the soft mode. 
The set of parameters for BaTiO$_3$ has been obtained using density 
functional theory simulation 
and is listed in 
Ref.~\onlinecite{Nishimatsu_et_al_2010}. 

Molecular dynamics simulations are performed by employing 
the feram code~\cite{feram_url}  developed by Nishimatsu 
\textit{et al}.~\cite{Nishimatsu_et_al_2008}  
The Nos\'e-Poincar\'e thermostat~\cite{Bond_et_al_1999} is used to 
simulate annealing starting well apart from the 
transition 
temperatures of our model.
At each $T$, after 120.000\,fs thermalization, the local 
dipoles ${\bm P}_i$ are averaged over 160.000\,fs and the 
final configuration is used to restart the next $T$ step.
We use  temperature sweep rates of up to $2 \times 10^{5}$~K/s, 
and 5\,K steps. 
At each $T$, the energy and the local 
polarization ${\bm P}_i$ allow us to determine the phase. 
Although the symmetry of the sample may be reduced by domains, 
we  speak of T and O phase if the system apart from domain walls 
has local polarization $P_{\left<100\right>}$ or 
$P_{\left<110\right>}$, respectively. 
We also use T and O to characterize phases with small 
field-induced monoclinic distortions, e.g. 
$0<P_{[010]}<<P_{[001]}$ is called T. 

In experiments, nucleation of different domains is observed at 
phase boundaries which can also be found in our simulations. 
However recall that in simulations, the material is under 
``\emph{ideal}'' conditions, i.e., absence of inhomogeneities and 
depolarization effects as well as there are finite size effects, 
therefore the probability of nucleation of domain walls is much 
smaller than that in experiments. Further we want to understand  
the effects of different types of domain walls separately which would 
not be possible for randomly nucleated structures. Therefore, we initialize our system 
with a well-defined local dipole arrangement at a temperature well 
away from the phase boundaries. This allows us to study properties of 
a desired ``geometry'' i.e.\ orientation and size of domains for a 
multi-domain phase without any additional effects. 
For this purpose we use a supercell size of $96 \times 96 \times 96$ 
formula units being equivalent to a sample of about $38 \times 38 
\times 38$\,nm$^3$ with periodic boundary conditions. 
We initialize equally spaced domain walls with distances of $d =$ 
19.1\,nm, 9.6\,nm and 4.8\,nm which correspond to presence of 2, 4 
and 8 DW in the supercell, respectively. 
For 90$^{\circ}$ walls in the O phase, we have furthermore considered 
$d=23.9$\,nm and $d=12$\,nm  using 2 and 4 domains 
in combination with  a $96 \times 96 \times 120$ supercell. For 
90$^{\circ}$ walls in the T phase, we restrict our study to the minimal number of 
4 walls 
with $d=13.5$\,nm compatible with the periodic boundary conditions for these walls along [110]. The number of domain walls considered here for 
each structure is determined by the periodic boundary conditions 
and the size of the supercell. 

\begin{figure}[h]
\centering
  \includegraphics[height=0.35\textwidth,clip,trim=1cm 1cm 2cm 0.8cm,angle=270]{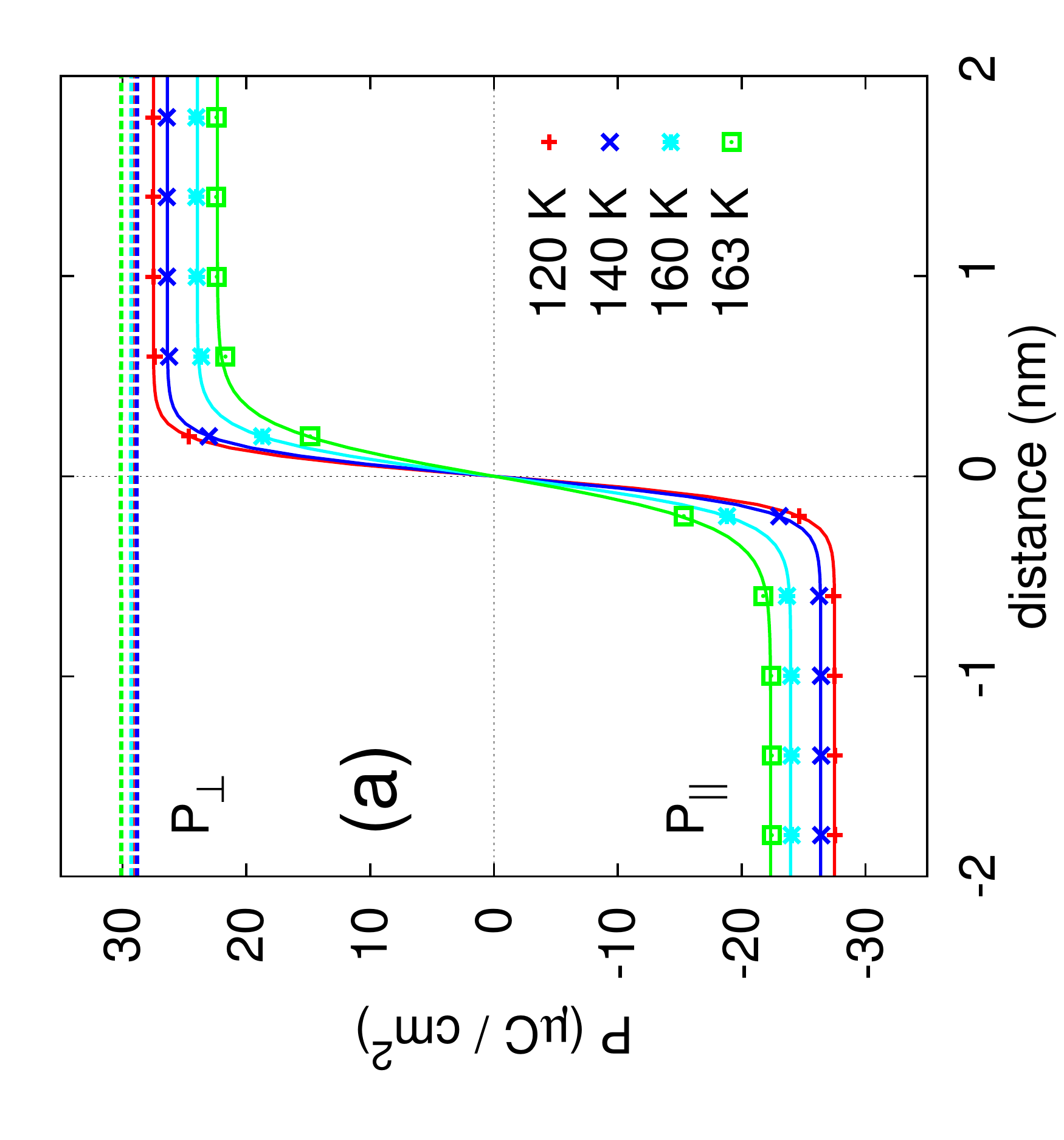}
  \includegraphics[height=0.36\textwidth,clip,trim=1cm 1cm 1.9cm 0.5cm,angle=270]{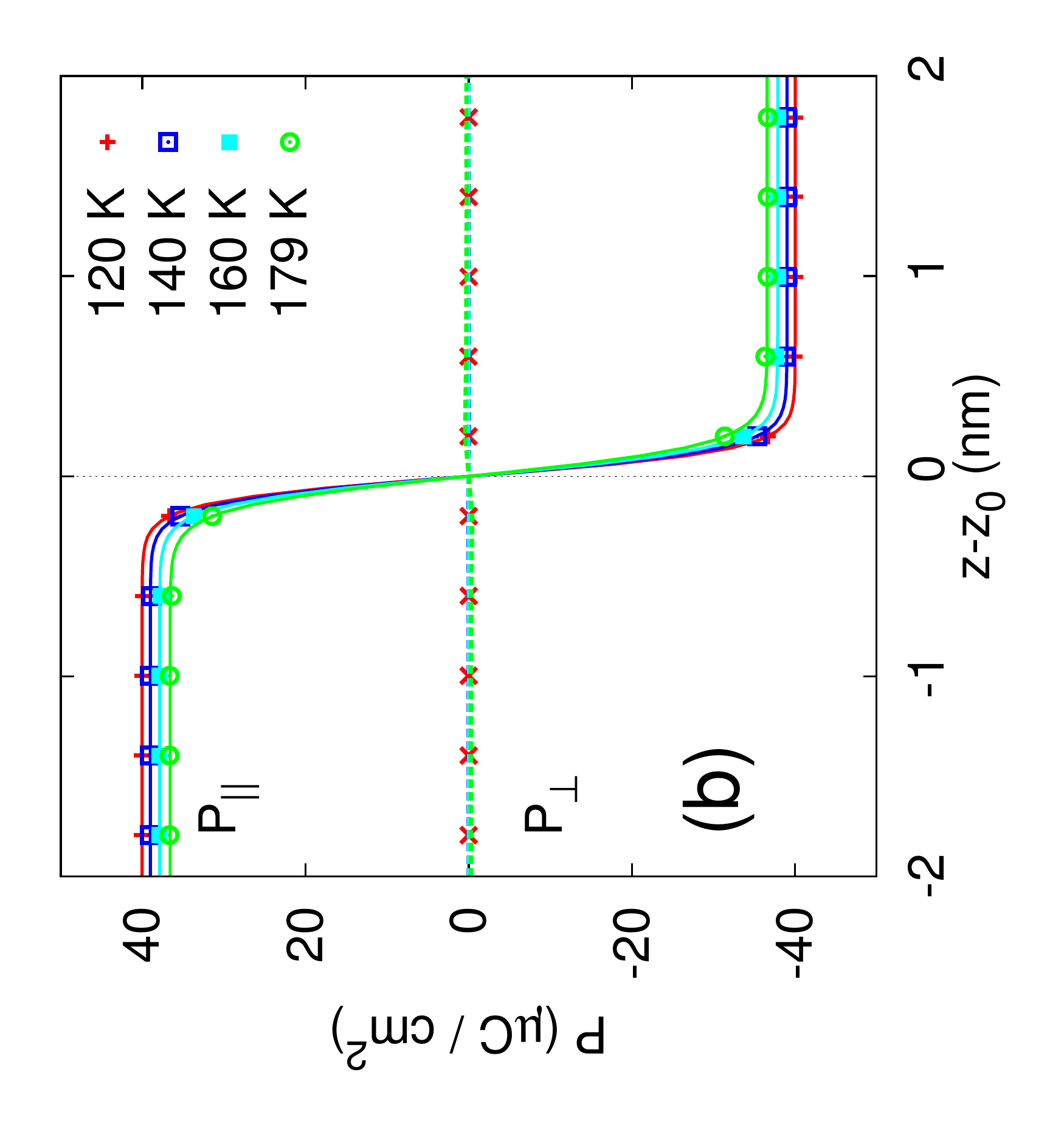}
    \includegraphics[height=0.37\textwidth,clip,trim=0.5cm 1cm 1cm 0.5cm,angle=270]{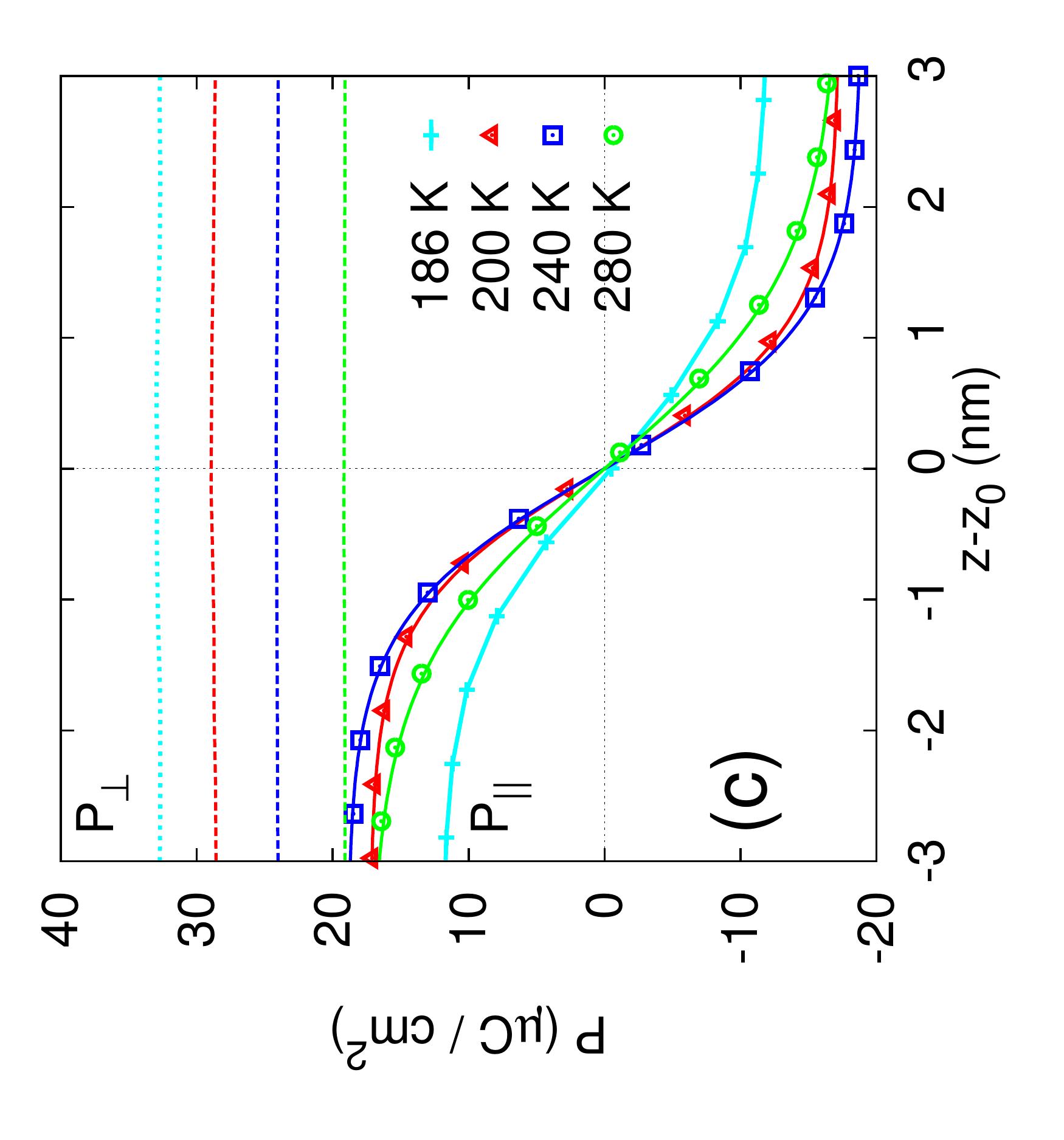}
	\caption{ 
	Average polarization per layer $\langle P_i \rangle_c (z-z_0)$ across O90 (a) 
	O180 (b) 
	and (c) T90 domain walls. Symbols
	correspond to simulated $P_{||}$, solid lines to tanh-fit 
	(Eq.~\eqref{eq:fit}) and dashed lines to $P_{\perp}$. 
	Here $z_0$ corresponds to the center of the DW. 
     We note that $T_C$ for configuration (a), (b), and (c) correspond to 
     163\,K and 179\,K and 184\,K respectively (discussed later in detail).
     }
\label{fig:indomain}
\end{figure}

\section{Results and discussion}
\label{sec:results}
As a starting point, we characterize multi-domain configurations 
in both tetragonal and orthorhombic phases separately using 
supercells with 
well-defined domain walls distances as described in 
Sec.~\ref{sec:comp}.\footnote{Note that the multi-domain states are only 
(meta-)stable above a critical minimal 
distance and hence systems with denser walls transform to the 
corresponding SD state at specific temperatures $T_t$; these systems 
are excluded from the introductory discussion. }
We project the polarization on the two components 
(i) parallel ($P_{||}$) and  (ii) perpendicular ($P_{\perp}$) 
to the wall;  these directions are defined 
for each configuration in Fig.~\ref{fig:domains}. 
We plot these two components for various MD configurations 
in Fig.~\ref{fig:indomain}.  
For the studied charge neutral walls, $P_{\perp}$ is constant across 
the domain walls as illustrated by dashed lines, 
with $P_{\perp}=P_{[001]}$ at the 
O90$^{\circ}$ wall, see (a), zero across non-elastic walls 
such as O180$^{\circ}$ as shown (b), and $P_{\perp}=P_{[110]}$ 
at the T90$^{\circ}$ wall, see (c). 
We do not find any significant change in $P_{\perp}$ for O90 and O180 walls, 
however for T90 walls, it increases as temperature is reduced towards 
the transition temperature.
The parallel component $P_{||}$ reverses the sign across the DW 
for all the configurations. This is expected based on underlying 
symmetries for the MD phases. 
$P_{||}$ approximately follows a tanh-profile as previously 
discussed in literature.~\cite{Marton_et_al_2010, Hlinka_Marton_2006,Grunebohm_Gruner_Entel_2012,Meyer_Vanderbilt_2002} 
and decreases if the system approaches $T_C$ in the presence of elastic walls.
We estimate the domain wall widths $2d_{\text{DW}}$ by fitting the 
polarization profile across a single wall with
\begin{equation}
P_{||}(z)=P_{0,||}(z_0)\cdot \tanh\left[\frac{(z-z_0)}{ d_{\text{DW}}}\right],
\label{eq:fit}
\end{equation} 
as shown with solid lines in Fig.~\ref{fig:indomain}.
Here $z-z_0$ corresponds to the distance from the center of the wall. 
\begin{figure}[tb]
    \centering
     \includegraphics[height=0.3\textwidth,clip,trim=2cm 3cm 10cm 6.5cm]{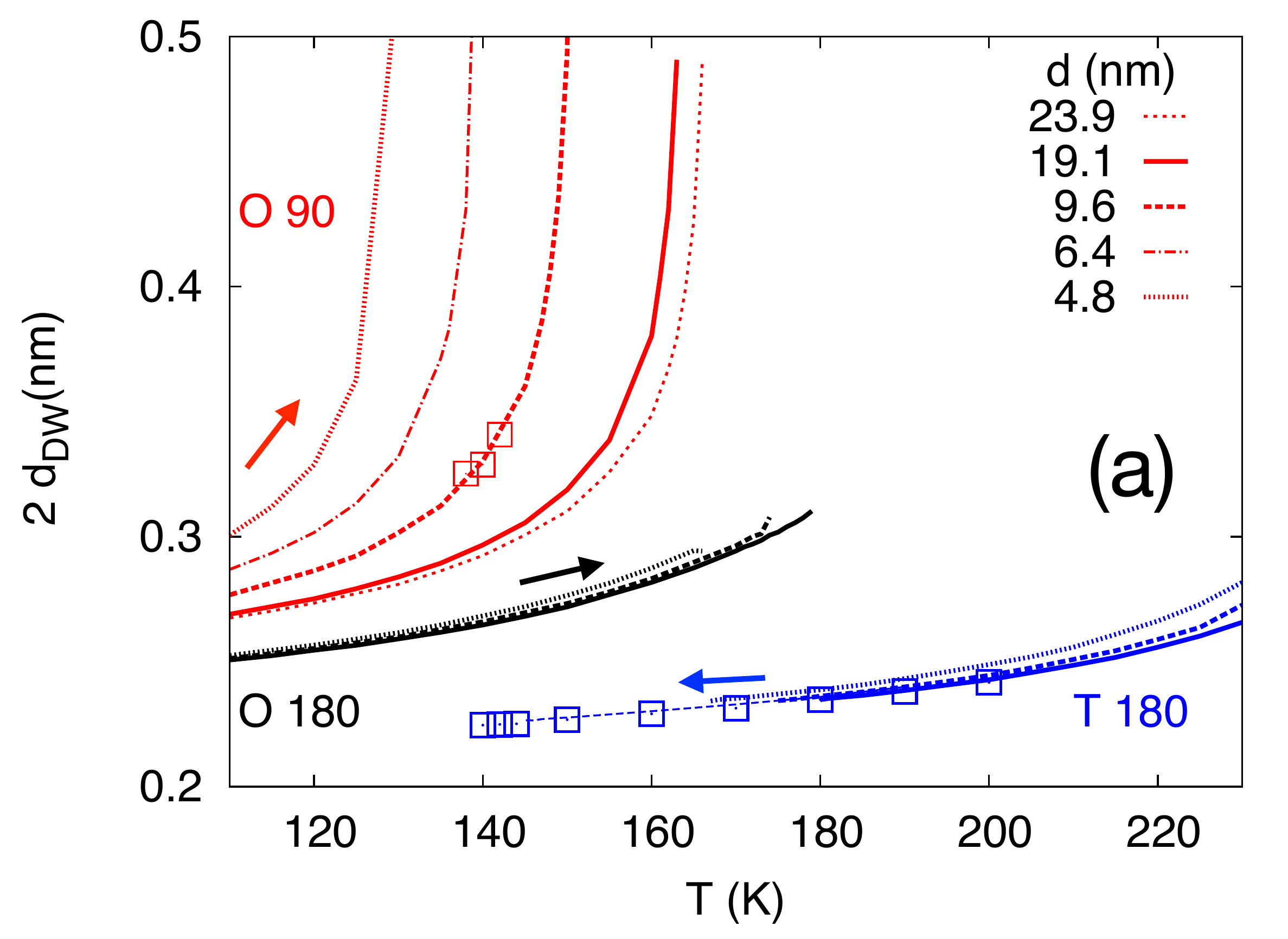}
    \includegraphics[height=0.3\textwidth,clip,trim=2cm 3cm 10cm 7cm]{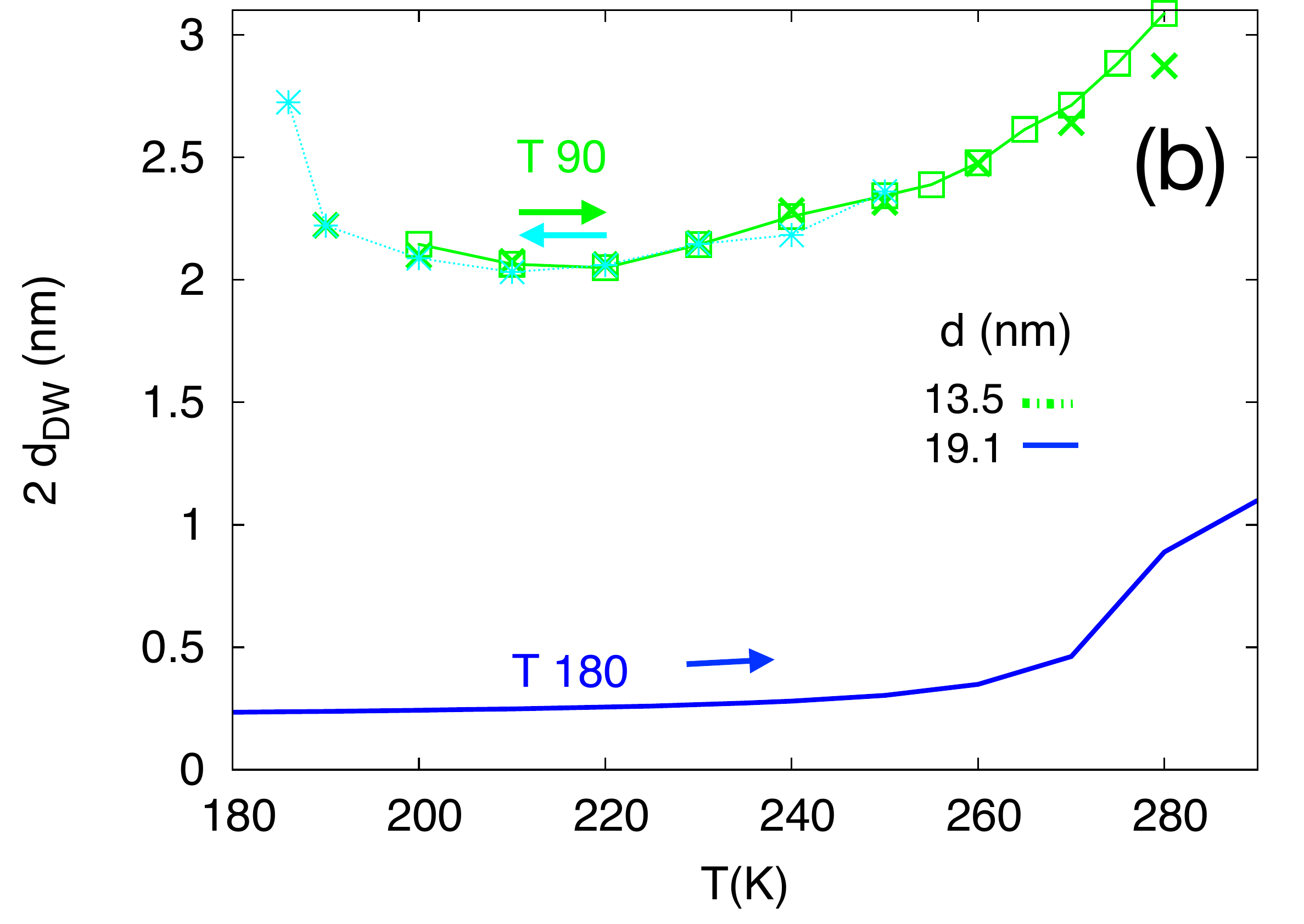}
    \caption{Temperature dependence of $2d_{\text{DW}}$ for
	O90 (red), O180 (black), T180 (blue) and T90 (green) walls (a) approaching the T-O transition (b) in the whole T phase. 
	Line types  mark the wall distances $d$ and the data has been recorded in heating simulations. For three examples also results for cooling simulations have been added (symbols). 
	}
    \label{fig:dDW}
\end{figure}
\begin{figure*}[bt]
\centering
\includegraphics[width=0.85\textwidth,clip,trim=1cm 16cm 1cm 1cm]{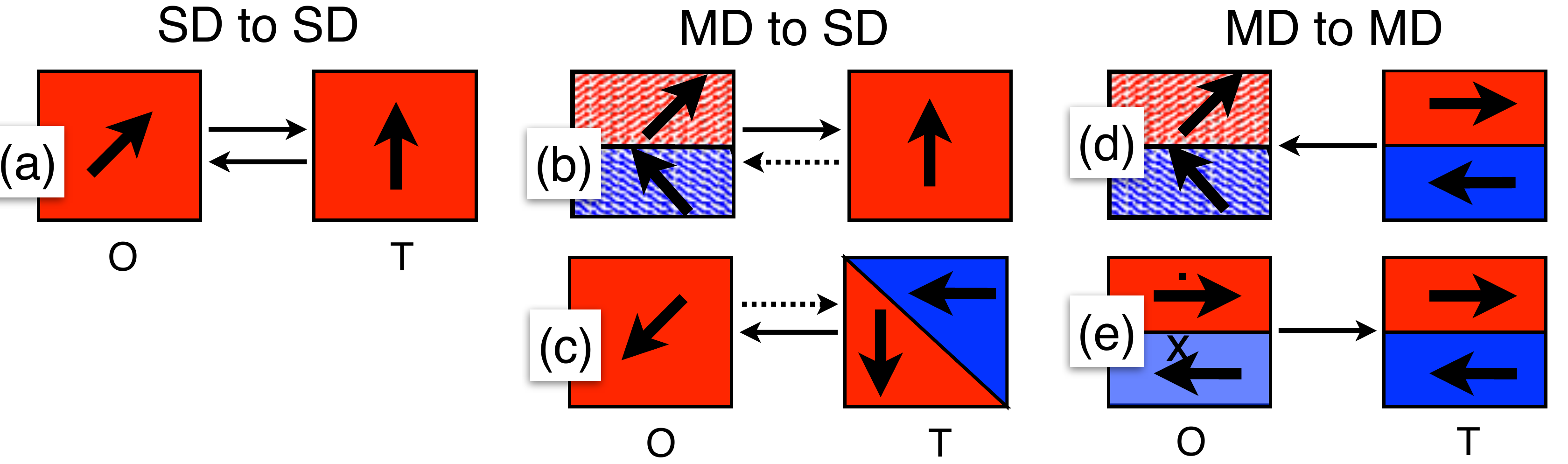}
    \caption{Schematic pictures of the observed changes of domain 
    structure at the T-O transition: 
    (a) SD O $\leftrightarrows$ SD T, e.g. between bc and c, 
    (b) O90 $\leftrightarrows$ SD T, e.g., bc/-bc to c,
    (c) T90  $\leftrightarrows$ O SD, e.g. between b/c and bc,
    (d) T 180$^{\circ}$ $\rightarrow$ O90$^{\circ}$, 
    e.g., from b/-b to bc/-bc, and 
    (e) O180 $\rightarrow$ T180, e.g. ab/-a-b to b/-b. 
    Note that in (e) dots and crosses mark polarization out of/in 
    the plane, respectively. Arrows between two phases 
    indicate transitions which are observed commonly (solid) and 
    only occasionally due to nucleation (dotted).
 \label{fig:transition} 
}
\end{figure*}

The calculated DW width is plotted in Fig.~\ref{fig:dDW} for different types 
of walls. We find that T90 walls are thickest 
followed by O90 and O180 walls and 
the T180 walls are the narrowest among those under study.   
This trend compares well with those results obtained from the 
Ginzburg-Landau-Devonshire model~\cite{Marton_et_al_2010} 
and density functional theory.~\cite{Grunebohm_Gruner_Entel_2012} 
Note that no quantitative agreement may be expected due to the different approximations in the three simulation methods. 
We reproduce the findings in Ref.~\onlinecite{shin_nucleation_2007} 
that the walls occasionally shift via a transient state 
with broadened domain walls.   
Further, we observe the predicted increase of $d_{\text{DW}}$ with 
increasing temperature.\cite{Marton_et_al_2010} 
For 180$^{\circ}$ walls, the temperature dependence of $d_{\text{dW}}$ 
is weaker compared to 90$^{\circ}$ walls and a large broadening 
occurs near the transition temperature for the latter. 
The broadening of the walls is related to fluctuations of the dipoles in 
the domain walls which strongly increase close to the transition temperature. 
Moreover we observe a considerable broadening of T180 and T90 
walls at high temperatures as the system approaches the 
ferroelectric to paraelectric transition temperature due to 
large fluctuations associated with this transition. 
The observation of domain wall broadening close to $T_C$ goes beyond 
predictions by Ginzburg-Landau theory~\cite{Marton_et_al_2010} probably 
 because the fluctuations of the order parameter 
are not fully included in that method. 

The domain density does not affect the DW width even for minimal 
distances of 4.8\,nm. 
It is further important to note that we obtain the same widths during the heating and 
cooling simulations as illustrated by symbols for some exemplary 
configurations in  Fig.~\ref{fig:dDW}. 
Thus this property is a material property at any temperature and is not 
influenced by thermal history or changes of domain wall positions, etc.
Along with this analysis of local polarization and DW widths, we 
also examined the energies of the MD configurations for different wall densities
which qualitatively reproduces the ordering of total 
energies published earlier.~\cite{Marton_et_al_2010} Furthermore, the domain wall profiles and energies  are converged with respect to the system size.
 From this analysis, we conclude that the MD configurations 
are well described within our model and simulation cell size.

We now turn to the main focus of our work: how do the domains 
evolve near the transition and what is their impact on the transitions? 
In absence of domains and external field, we observe a sharp first-order 
transition with a 
jump of order parameter and energy at the transition temperature. 
Our calculated transition temperatures are $T_C^c = $ 138\,K for cooling 
and $T_C^h = $ 186\,K for heating, that is, we get a thermal hysteresis 
of about 50\,K.
The quantitative disagreement with experimental data is expected as the used model systematically underestimates  transition temperatures and as the thermal hysteresis is considerably higher in ideal 
material.~\cite{Marathe_et_al_2017}

In the presence of domains, we observe different types of transitions 
between SD and MD phases which are summarized in 
Fig.~\ref{fig:transition}. In all cases, the walls do not disperse during the phase transition and the local polarization 
rotates by 45$^{\circ}$ analogous to the SD 
case shown in Fig.~\ref{fig:transition}(a). 
This results in two different scenarios -- 
(1) for 90$^{\circ}$ walls, the 
polarization in both domains rotates to the initial $P_{\perp}$ 
direction and transforms to the SD phase (panels b--c); and 
(2) for 180$^{\circ}$ walls, no such rotation is possible and we observe 
T180 to O90 and O180 to T180 transitions (see panels d--e). 
We note that the first-order character of this 
phase transition persists in the presence of domains walls, 
and we observe an abrupt transition with a jump in energy and 
polarization for all the systems considered. 
Occasionally, multi-domain phases nucleate near the transition 
even for initial SD phases as illustrated by dashed arrows in 
Fig.~\ref{fig:transition}. 
However, we do not discuss these transitions here as the probability 
of such nucleation is rather low in our simulations.

\begin{figure}[t]
   \centering
   \includegraphics*[width=0.38\textwidth,clip,trim=3cm 8.5cm 14cm 0.5cm]{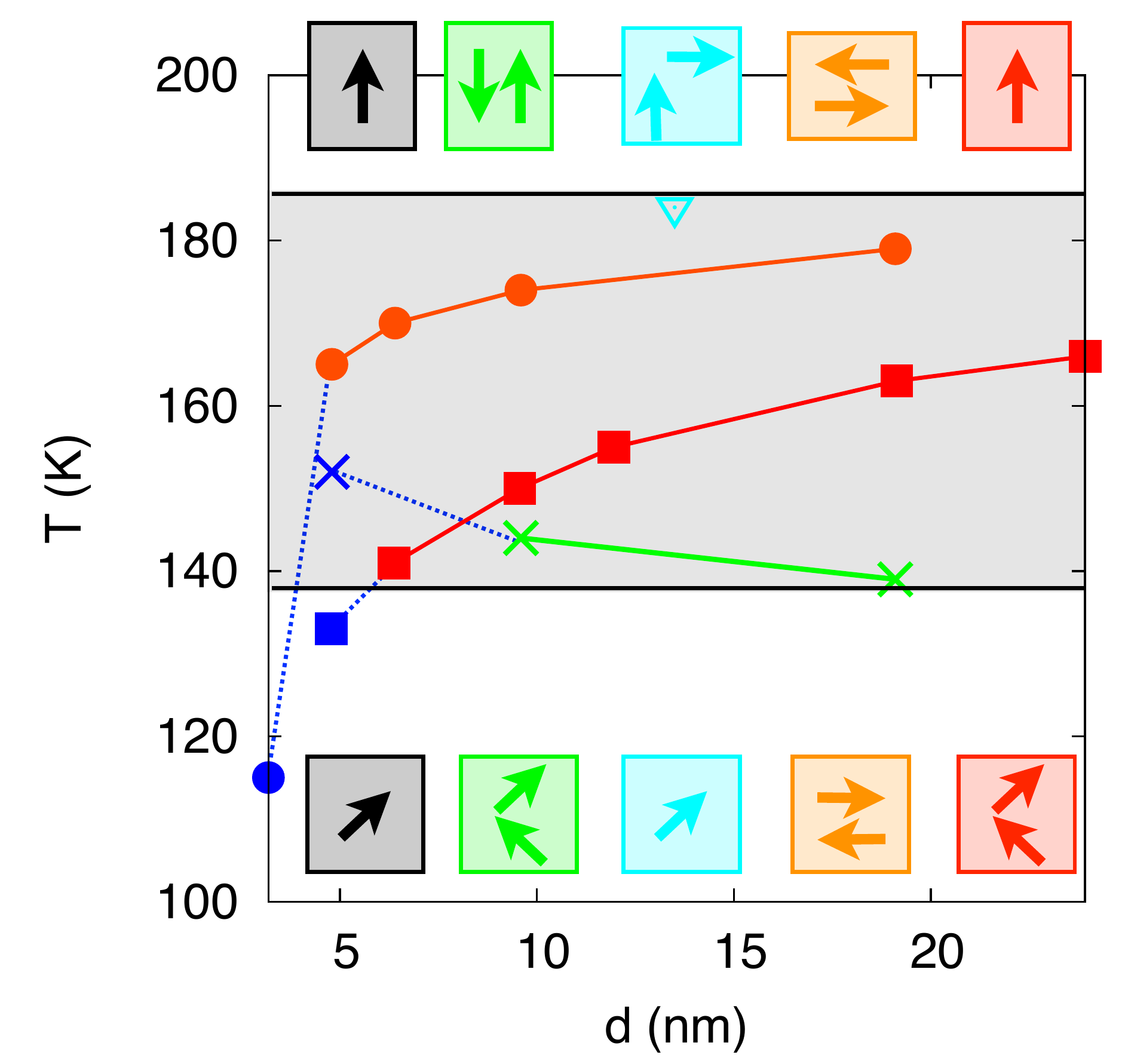}
   \caption{ Transition temperatures without external field as a 
   function of the domain wall distance for transitions   
   O90 $\rightarrow$ T-SD (red squares), 
   T180 $\rightarrow$ O90 (green crosses),  
   O180 $\rightarrow$ T180 (orange circles) and 
   T90 $\rightarrow$ O-SD (cyan triangle).  
   These transitions are illustrated in the insets. 
   For dense walls, blue symbols mark temperatures $T_t$ 
   at which the MD configuration transforms to the ground state 
   SD configuration without any phase transition. 
   The gray shaded area illustrate the coexistence range for 
   T-SD to O-SD transition. 
    \label{fig:TC_d}
    }
\end{figure}

Transition temperatures in the presence of domain walls are plotted 
as a function of wall densities in Fig.~\ref{fig:TC_d}.
\footnote{
 Note that domain walls occasionally move and the domain wall distances 
 are no longer equidistant. However, for simplicity, we 
 name the MD configurations following the initial domain distances. 
 }
Domains generally promote the phase transition and thus reduce 
$T_C^{h}$ and enhance $T_C^{c}$ as compared to the SD phase. 
For 180 walls, $T_C$ converges slowly towards the SD phase boundaries 
with increasing wall distances therefore when starting with O-MD phases 
$T_c^h$ increases with $d$ while for T-MD phases $T_c^c$ decreases.  The convergence towards the SD phase boundary is generally expected 
because with decreasing DW density (equivalently increasing $d$) 
the volume fraction of dipoles near a DW as well as potential interactions between walls decrease. 
For O90 walls, the change of $T_C$ with the wall distances is considerably smaller.
Note that for the densest DW considered within our study, 
the MD phases are not stable in the vicinity of the phase transition 
and the system transforms to the 
SD phase (that is, O-MD to O-SD and T-MD to T-SD phases) at 
a temperature $T_t$; indicated in blue in Fig.~\ref{fig:TC_d}. 

The change in the transition temperatures is larger for 90 walls 
than for 180 walls. For example, heating from the O phase at 
$d = 9.6$\,nm, we observe $T_C^h = 174$\,K for O180 
and 151\,K for O90 walls, respectively.  
While cooling down from the T-MD phases, the increase of $T_C^c$ 
by T180 walls is small (green crosses), whereas we observe a reduction 
of the thermal hysteresis to only 2\,K in the presence of T90 walls 
with $d = 13.5$\,nm.~\footnote{Recall that unfortunately due to 
computational restrictions we have not performed a systematic study 
for this type of wall.} 

\begin{figure}[tb]
    \centering
    \includegraphics[width=0.5\textwidth,clip,trim=5cm 7cm 12cm 3cm]{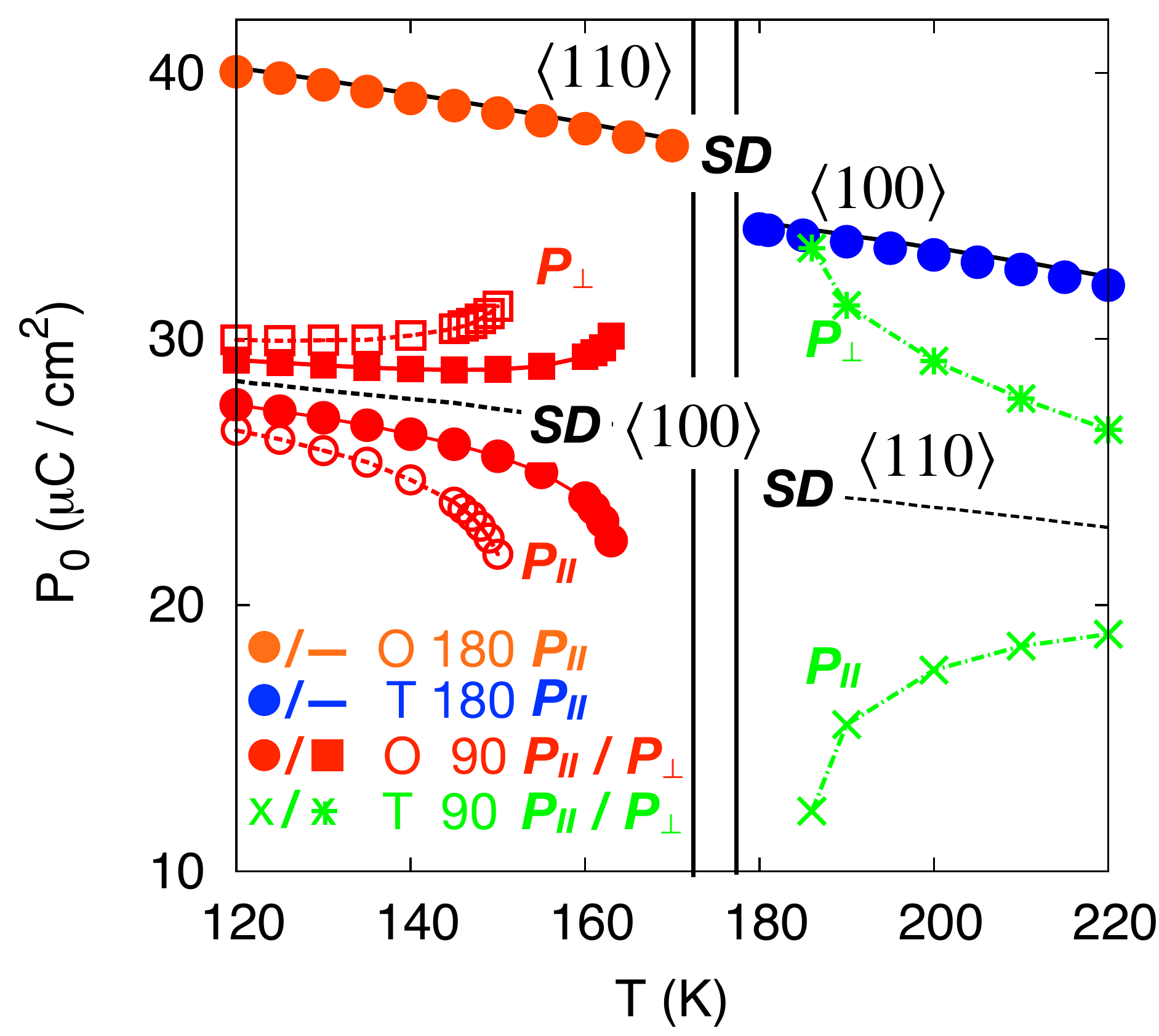}
	\caption{
	Projections of polarization parallel and perpendicular to the 
	wall ($P_{||,0}$ and $P_{\perp,0}$) in the center of -- 
	O90 (red) for $d =$ 19.1\,nm (solid lines) and 9.6\,nm (dashed lines); 
	O180 (orange) for $d =$ 19.1 \,nm; 
	T180 (blue) for $d =$ 10.1 \,nm and T90 (green) for $d =$ 13.5\,nm 
	configurations. Here we excluded regions near 
	phase transitions.  $P_{||,0}$ has been determined 
	by Eq.~\eqref{eq:fit} and  $P_{\perp,0}$ corresponds to the 
	sample average. As a reference, 
	SD results projected on the same directions are added in black. 
 }
    \label{fig:P0_DW}
\end{figure}

To analyze the impact of domains on $T_C$, one may distinguish between 
rotation of polarization and strain variation observed 
in the domains (``volume effect'') and  the properties of domain walls 
themselves (``interface effect"). 
To examine the volume effect, we look at the changes in polarization 
at the center of the domains where the effects arising from the interfaces themselves, 
i.e.\ the domain walls, should be minimal. 
For this purpose, we calculate the polarization in the center of the domain 
$P_{||,0}$ using Eqn.~\eqref{eq:fit}. The resulting polarization in the 
domain centers is plotted in Fig.~\ref{fig:P0_DW} as a function of temperature. 
In case of 180$^{\circ}$ walls, the $P_{||}$ component matches exactly 
to the magnitude of the overall polarization of the SD phases and $P_{\perp}$ 
is equal to zero. The impact of temperature on the 
polarization direction in both domains is negligible. 
Thus, 180$^{\circ}$ DW have no impact on the local polarization in the adjacent domains. 
In case of O90$^{\circ}$ walls, $P_{\perp}$ and $P_{||}$ both correspond to
$\langle 100 \rangle$ directions. Thus, if domain walls had 
no impact on the local polarization in the domain center, one would expect 
$P_{\perp}=P_{||}$ as $P=P_{\perp}+P_{||}$ would point along the
$\langle [110] \rangle$ direction of a particular orthorhombic variant.
In contrast to that we observe a monoclinic distortion of the polarization 
with  $P_{\perp,0}>P_{||,0}$, cf. red squares and red circles in 
Fig.~\ref{fig:P0_DW}. 
The  monoclinic distortion is even larger in case of T90 wall for which 
the $P_{\perp}$ and $P_{||}$ both correspond to 
$\langle 110 \rangle$ directions, cf.\ green stars and crosses.
Away from phase transitions, the monoclinc distortion increases with the 
domain wall density. 
The considerable monoclinic distortion even in the center of domains 
shows that the the elastic walls induce long-range modification 
in the system both in polarization and strain. 
This explains much slower convergence of $T_C$ with 
respect to $d$ for O90 walls compared to other configurations. 

Approaching the transition temperatures, i.e., heating O90 or 
cooling T90 configurations, the monoclinic distortions increase 
for MD configurations.  
Enhanced monoclinic distortions close to ferroelectric-ferroelectric 
can be related to reduced anisotropy energy 
and energy penalty of reorientation of dipoles near the phase 
boundary and allow for monoclinic bridging phases which have been 
discussed earlier.~\cite{Rossetti_2015,acosta_origin_2015} 
Such a monoclinic bridging state generally reduces the energy barrier between 
different phases promoting the phase transition. 
The changes in $T_C$ in our study are related to the size of the monoclinic distortion and indeed the largest change observed for 
T90 configuration corresponds to the largest distortion obtained
in our study. 

Secondly to examine the interface effects, we look at the 
properties of domains walls themselves. 
As discussed earlier (Fig.~\ref{fig:dDW} and subsequent discussion), 
the 180$^{\circ}$ walls are extremely thin and their width doesn't 
diverge even in the vicinity of the phase transition, whereas, 
the elastic walls are slightly wider  and diverge as the system 
approaches transition. Furthermore, 180$^{\circ}$ walls are 
Ising-like, i.e., the polarization in the center of the wall 
does not point along the polarization direction of either domains 
but rather decreases in magnitude, only.  
However, for the O90 wall, the polarization in the center of the wall 
is pointing along $\langle 100 \rangle$ 
(i.e., along the spontaneous polarization of the T-phase), 
and for T90 wall, along $\langle 110 \rangle$ 
(i.e., along the spontaneous polarization of the O-phase) 
and may thus act as nucleation centers for the transition. 
Thus for 90$^{\circ}$ walls, both volume and interface effects come 
together to accelerate the phase transition, whereas for 
180$^{\circ}$ walls, only interface effects contribute to the 
change in $T_C$. 

Interestingly, the same trends also occur at the 
orthorhombic to rhombohedral transition where both O90 and O180 
phases transform to R109 configurations with $[\pm1\pm11]$ domains 
and with $P_{||}<P_{\perp}$ similar to the O90 case. 
Similar to the discussion on the T-O transition, $T_C$ is enhanced 
by only 5--10\,K and at the same time, there is no  considerable 
monoclinic distortions in the inital O phase. 
This points to the general trend that the long-range monoclinic distortions induced by elastic walls are important to modify $T_C$. 
However, a detailed study of this transition is out of scope 
for this paper.

\begin{figure}[!h]
\begin{center}
   \includegraphics[width=0.45\textwidth,clip,trim=3cm 11cm 8.1cm 1.5cm]{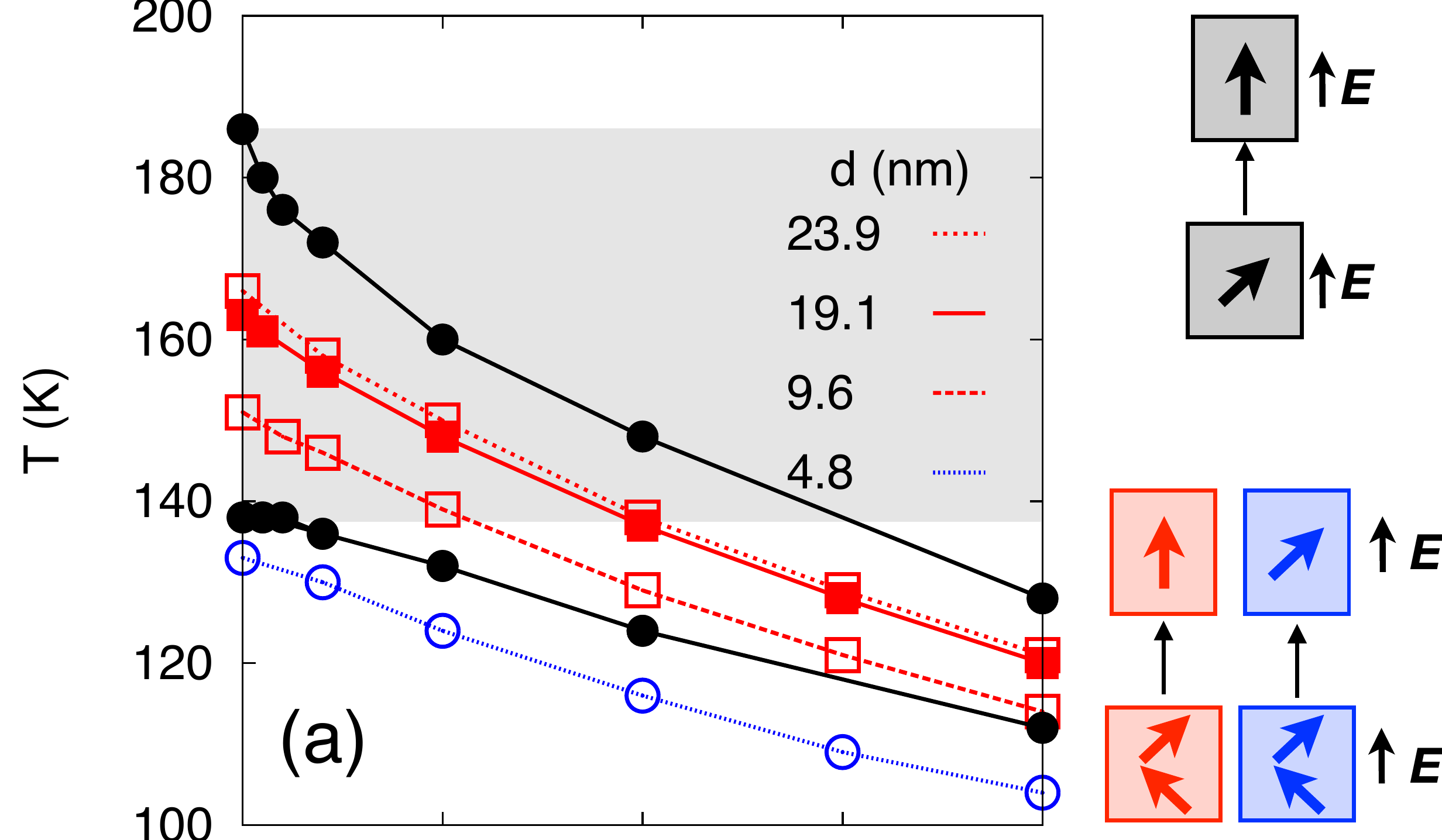}
   \includegraphics[width=0.45\textwidth,clip,trim=3cm 11cm 8.1cm 1cm]{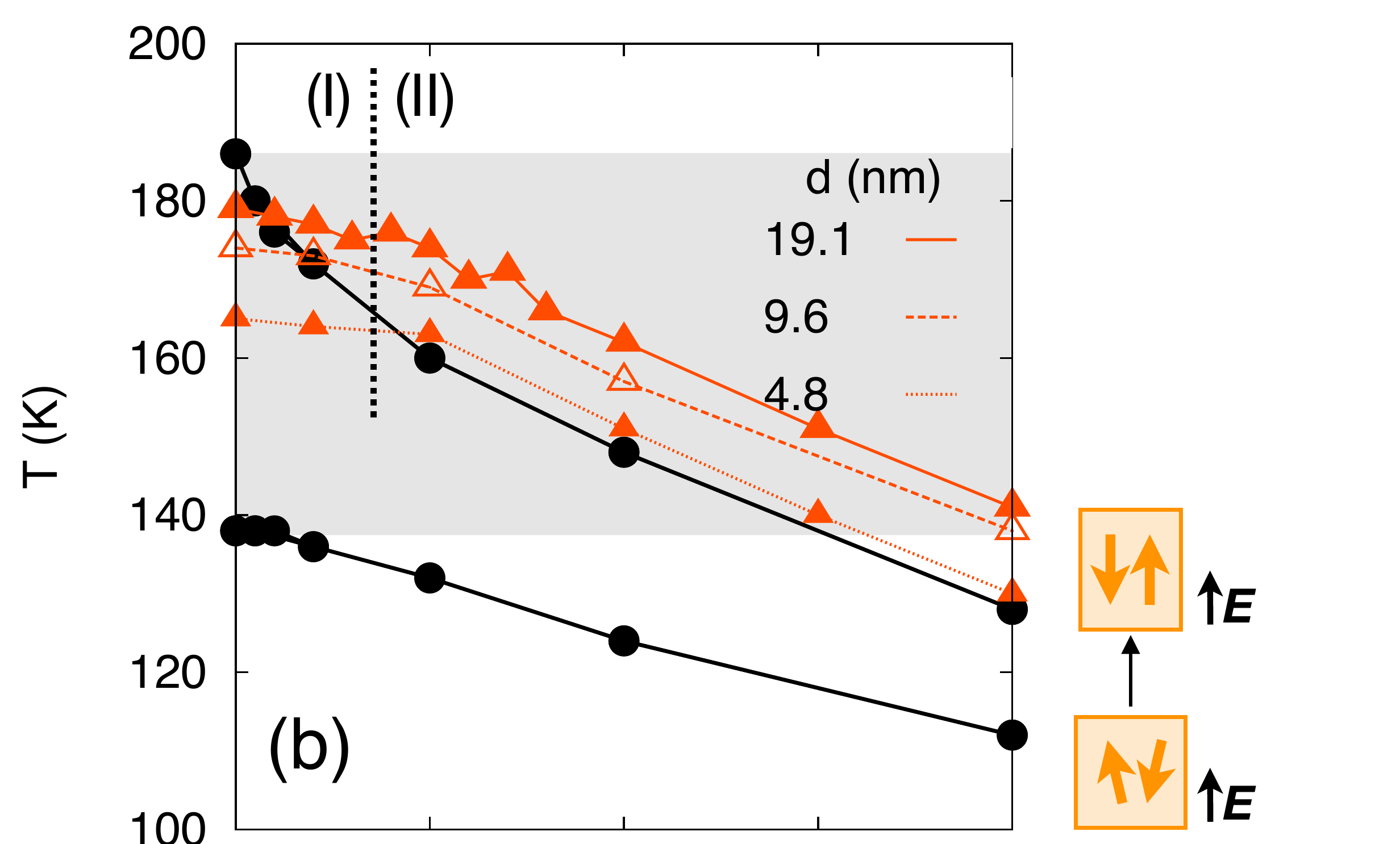}
   \includegraphics[width=0.45\textwidth,clip,trim=3cm 8.5cm 8.1cm 1cm]{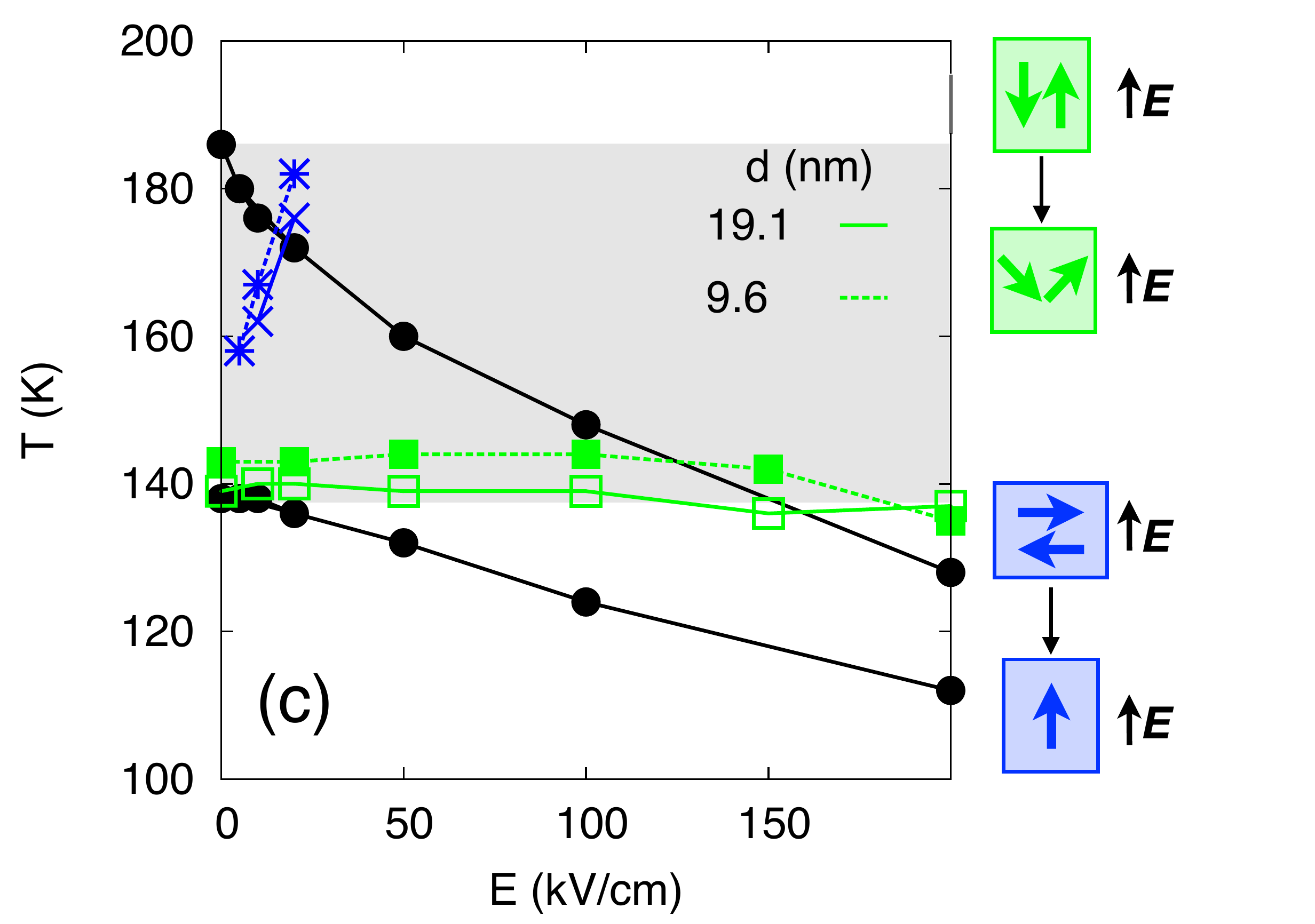}
     \caption{
     Change of T $\leftrightarrow$ O transition temperatures 
     in the presence of an external field applied along $P_{||}$ 
     (unless mentioned otherwise) 
     for (a) O90, (b) O180, and (c)  T180 walls. 
     Gray shaded areas and black lines mark the coexistence region of 
     the SD reference configuration without and with field taken from
     Ref.~\onlinecite{Marathe_et_al_2017}, respectively. 
     Line types refer to different domain distances (see legends) and 
     colors characterize the phase transitions as follows: 
     red: O90 to T SD; orange: O180 to T180; green: T180 to O90;  
     whereas in blue, $T_t$ is shown (O90 to O-SD, 
     T180 to T-SD). Each type of transition is shown schematically 
     in the insets.  In (b), vertical dashed line marks field range (I) small slope 
    of $T_C(E)$ and (II) larger $T_C(E)$. In (c), blue stars and 
    crosses indicate the transition 
     temperatures when the field is applied perpendicular to the 
     wall. 
  }
  \label{fig:TC}
  \end{center}
\end{figure}

Next, to shed light on the DW-field coupling, we performed cooling and heating simulations in the presence 
of an external field  along $\langle 100\rangle$. 
Recall that for MD phases, the relative orientation between the field and local polarization in each domain may be different, and the direction 
of the field relative to the DW also influences the field coupling. 
Unless specified otherwise, we applied the field parallel to the 
DW.~\footnote{{
Our chosen field direction $\langle 100 \rangle$ put certain limits 
on the studied cases. 
For example, for T90 walls, the domains wall is parallel 
to $\langle 110\rangle$ direction, therefore these walls are not included 
in this study of wall-field coupling. }}
The calculated transition temperatures are plotted in 
Fig.~\ref{fig:TC} as a function of applied field magnitude 
at several $d$.  
Note that for SD phases, both $T_C^c$ and $T_C^h$ reduce systematically 
with the increasing field magnitude. 
This can be understood because an external field along 
$\langle 001\rangle$ favors the T-phase with the induced polarization 
along the applied field direction as has been reported 
by us in an earlier study;~\cite{Marathe_et_al_2017} as a reference 
the data is reproduced here, see black lines in Fig.~\ref{fig:TC}. 
For MD phases, we observe three distinct trends: 
first, a comparable change in $T_c$ with $E$ to that found for SD phases 
for O90 walls (panel a) and O180 walls (region II in panel b);
second, no or a very small change in $T_c$ with increasing field strength 
for O180 walls (region I in panel b) and for T180 walls (panel c green 
squares); and 
third, the reduction in the stability range of dense O90 walls and T180 
walls with $E$ (blue symbols in panels a and c respectively), i.e. 
$T_t (E)$  
increases in cooling simulations and decreases in heating simulations. 
The shifts in $T_C$'s with respect to $d$ which already have been 
discussed in Fig.~\ref{fig:TC_d} are not affected 
by an application of the field. 

\begin{figure}[tb]
    \centering
    \includegraphics[width=0.45\textwidth,clip,trim=6cm 6cm 12cm 3.5cm]{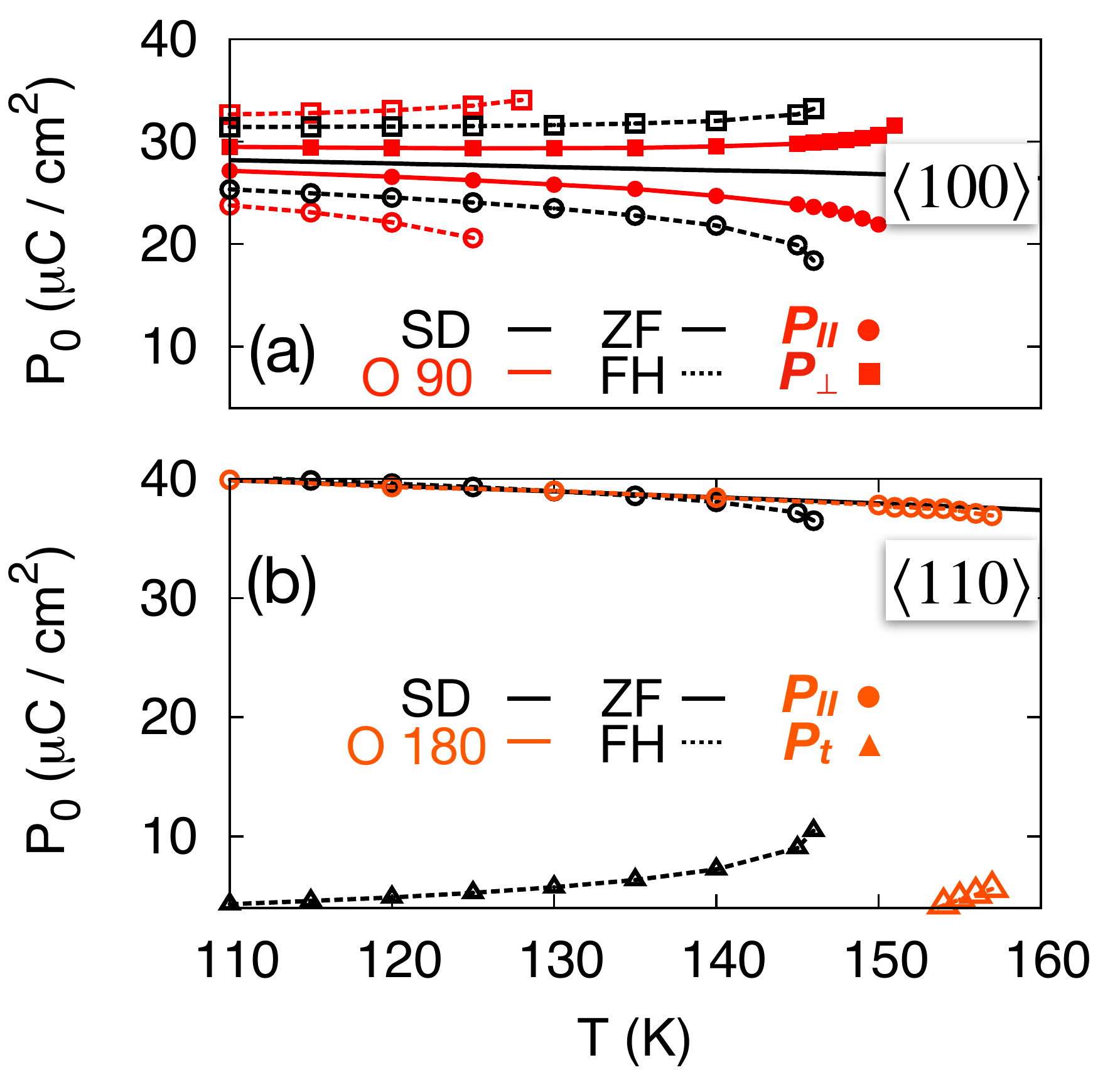}
    \caption{
    Monoclinic distortions in domain centers for a field of 100~kV/cm
    in the presence of O90 (a) and O180 (b) walls (with $d =$ 9.6\,nm). 
    For reference, the same projections are added for SD phase (black) 
    and zero field (solid lines). The field induces larger monoclinic 
    distortions for O90 walls ($P_{\perp}$ (squares) $>$  $P_{||}$ 
    (circles)) compared  to both zero field phase as well as SD phase 
    under the same field. For O180 walls, the field induces a 
    polarization component $P_t$ (triangles) along the direction 
    which is orthogonal to both $P_{||}$ and $P_{\perp}$ only for 
    large applied fields and this distortion is smaller compared 
    to the SD phase. 
    \label{fig:rot_double_field}
    }
\end{figure}

Let us now look at the $T_C$ trends for MD configurations in more 
detail. In the first case, for O90 walls, all domains have the 
polarization component $P_{\perp}$ along the applied field direction which 
results in the partial rotation of dipoles towards the field 
direction throughout the system and finally the SD T phase 
parallel to the field. 
 Recall that O90 walls promote the transition by DW broadening and 
 an induced monoclinic symmetry throughout the domains.
The applied field has a similar impact on the polarization direction 
and further enhances the  monoclinic distortion with $P_a>P_b$ for 
the [100] field which is rather temperature independent at low 
temperatures and increases considerably close to $T_C$,  
see Fig.~\ref{fig:rot_double_field}(a). 
The field also broadens the DW (not shown); 
these two effects add up and $T_c$ is reduced with increasing field.  
For the densest walls ($d =$ 4.8\,nm), the system 
is meta-stable near the phase transition even in the absence of an 
external field.  The field-induced enhanced monoclinic distortions 
reduce $T_t$ systematically with field strength. 

The case of O180 walls transforming to  T180 is more complex and 
interesting because  
the field is neither fully aligned with initial or final 
polarization nor do all domains incline the same angle with $E$. 
For a field along [010], we observe two different scenarios: 
I. for small fields ($E \le$30\,kV/cm), the polarization along 
$\pm[110]$ transforms to the T180 phases with $\pm[100]$, i.e. 
perpendicular to the  field. In this case we neither observe a 
broadening of the domain wall nor a rotation of $P$ towards 
the field direction and $T_C$ shows a small dependence on 
applied field. 
II. for larger fields we observe a rotation of the dipoles in both 
 domains to a state with monoclinc distortion and $|P_b|>|P_a|$ 
 promoting the transition to the $\pm[010]$ phase collinear with 
 the external field.  
 Recall that without field O180 walls do not induce any monoclinic 
 distortion. 
In the small field scenario I, we even observe a reduction 
of the polarization 
component antiparallel to the field, i.e.\ for a field along [010], 
the -b component in the -a-b domain is reduced. 
Mediated by the strain the perpendicular tetragonal direction is thus stabilized. 
In this case, the field neither considerably reduces the energy barrier 
 for the phase transition nor does it stabilize the final state (as $P$ 
 in each domain is perpendicular to the field) and in turn $T_C$ is not 
 lowered by the field. 
  The  interesting consequence of this is that the transition temperatures 
  fall outside of the coexistence region of the SD phase with the field. 
In contrast, higher fields (case II) overcome the energy barrier for a 
monoclinic distortion close to $T_C$, see polarization component P$_t$ in 
Fig.~\ref{fig:rot_double_field}~b. Analogous to the SD case, the distortion 
promotes the phase transition and we obtain a similar slope of $Tc(E)$ 
in this field range. 
 
For T180 walls,  
the response for the field applied collinear to $P_{||}$ is similar to that 
observed for O180 walls with 
small fields and again can be understood by a reduced field coupling. 
  The applied field would favor the parallel SD tetragonal phase. 
  However, it is not strong enough to eliminate the  walls. Instead 
  the T180 domains with $[0\pm10]$ transform to O90 domains with 
  $[0\pm11]$. In both phases the field is parallel to $P_{||}$ in 
  one domain and would favor an increase of it, and by the strong 
  strain-polarization coupling,  also of the tetragonal strain. 
 The effect of the field is opposite in the other domain reducing 
 polarization and strain.
 Such a mismatch at the wall is unfavorable and thus the field-coupling is 
 compensated in both phases and $T_C$ is barely modified with field magnitude.
 
In this work, we focus on  only one relative direction between 
field and 
domain wall for each type of walls.
One would expect an anisotropic response to the applied field 
due to presence of DWs. The complete study of  the anisotropy of the 
field is out of scope in this paper.  As an example, 
we however applied $E_{\perp}$ to T180 configuration and the 
corresponding results are shown in blue in Fig.~\ref{fig:TC}~(c). 
Here, the applied field is perpendicular to polarization in both
domains. In this case the energy barrier for rotating the dipoles 
uniformly towards the field direction is small and further 
decreases as the system approaches $T_C$. 
Therefore, the walls easily disappear and the system 
transforms to the T-SD phase at $T_t$ which increases with the field 
magnitude. 
To sum up, we show that the field coupling depends crucially 
on the relative direction of field and local polarization 
in the domains and in particular the combination of elastic 
walls and electric field along $P_{\perp}$ leads to a
considerable reduction in thermal hysteresis and may thus 
allow for large reversible functional responses.

\section{Summary and Outlook}
\label{sec:conclude}
We conducted a systematic study of the coupling of phase transition, 
domain walls and electrical field based on molecular dynamics simulations 
of an {\it{ab initio}} effective Hamiltonian. 
Starting from the idealized bulk BaTiO$_3$ system, we introduced low-energy 
MD configurations in the orthorhombic and tetragonal phases which 
are commonly found as complex superstructures in experiments in order to separately discuss their impact on the phase transition. 

Thereby we could show that these MD states once nucleated or 
initialized in the system remain meta-stable within a 
temperature range of interest, down to domain width of 4.8\,nm. 
Our observations point  to the fact that the monoclinic symmetry 
observed close to the O-T 
transition in experiments~\cite{Eisenschmidt_et_al_2012} may be
related to elastic domain walls present in the system. 
Another important consequence of our findings is that the 
large spread observed in the experimental transition temperatures 
can be partly 
related to different domain structures in the sample with no, or
mainly 180$^{\circ}$ walls resulting in a broad hysteresis, while 
elastic walls may induce a monoclinic bridging phase. 

Our most exciting finding is that elastic domain walls strongly 
modify the transition temperatures and reduce the thermal hysteresis 
by one order of magnitude. On the other hand, the impact of 
180$^{\circ}$ walls is small. These trends are further amplified in the 
presence of an external field. The combination of external field and 
O90 walls allows to efficiently push $T_C$ below the zero-field coexistence 
range which is needed for a large reversible functional response. 
In contrast, 180$^{\circ}$ walls reduce the coupling between $T_C$ 
and the external field and as a result enhance the thermal hysteresis 
in the presence of an external field compared to the SD case. 
We have shown that these different trends can be attributed to 
I. volume effects, that is monoclinic distortions within domains, and 
II. interface effects, that is, large and diverging DW widths.

Our results are a significant finding for prospective applications 
using domain wall engineering to reduce thermal hysteresis of the 
ferroelectric to ferroelectric transition.
Elastic domain walls may act as a bias to reduce the field strength 
needed to induce a complete transition in the system and thus 
yielding large functional responses. 
This reduction of thermal hysteresis combined with the enhanced 
piezoelectric response obtained for these 
systems,~\cite{Wada_et_al_2005}  
make domain-wall engineered materials especially attractive for 
applications such as solid-state cooling 
which require large response near room-temperature. 
With advanced and precise techniques 
available to control the growth of 
nano-ferroelectrics,~\cite{Schilling_et_al_2009} we believe that
our work would provide a practical guide for further development
of functional devices. 

\section{Acknowledgements}
AG would like to acknowledge financial support by the German research 
foundation (GR 4792/1-2 and 4792/2) and computational resources provided 
by the Center for Computational Science and Simulation (CCSS), 
University of Duisburg-Essen. 
MM would like to acknowledge the funding from the European Union’s 
Horizon 2020 research and innovation programme under Marie 
Sk\l{}odowska-Curie grant agreement No 665919.

\bibliography{literature}

\end{document}